\documentclass[a4paper,fleqn,usenatbib]{mnras}
\pdfoutput=1
\usepackage{newtxtext}
\usepackage{mathptmx}

\usepackage[T1]{fontenc}
\usepackage{ae,aecompl}
\usepackage { float }  
\usepackage{xcolor}

\usepackage{graphicx}	
\usepackage{amsmath}	
\usepackage{amssymb}	
\usepackage{subcaption}
\usepackage{booktabs}
\usepackage{array}

\newcommand{\OIII}{[OIII]$\lambda$5007}
\newcommand{\OI}{[OI]$\lambda$6300}
\newcommand{\La}{Ly$\alpha$}
\newcommand{\Ha}{H$\alpha$}
\newcommand{\Hb}{H$\beta$}
\newcommand{\kms}{km s$^{-1}$}
\newcommand{\Mo}{M$_\odot$}
\newcommand{\x}{$\times$}

\newcommand{\OIHa}{[OI]$\lambda$6300/H$\alpha$}
\newcommand{\OIIIHa}{[OIII]$\lambda$5007/H$\alpha$}
\newcommand{\OIIIHb}{[OIII]$\lambda$5007/H$\beta$}

\title[Stellar feedback impact in Haro\,11]{The impact of Stellar feedback from velocity-dependent ionized gas maps. -- A MUSE view of Haro 11.
\thanks{Based on observations made with ESO Telescopes at the La Silla Paranal Observatory under programme IDs 094.B-0944(A) and 096.B-0923(A).}}

\author[V. Menacho et al.]{
V. Menacho$^{1}$, 
G. \"Ostlin $^{1}$,  
A. Bik$^{1}$,  
L. Della Bruna$^{1}$,
J. Melinder$^{1}$,  
A. Adamo$^{1}$, \newauthor \space
M. Hayes$^{1}$,
E.C. Herenz$^{1}$,
N. Bergvall$^{2}$
\\
$^{1}$Department of Astronomy, Stockholm University, AlbaNova, 106 91 Stockholm, Sweden\\
    \space  Oskar Klein Centre, AlbaNova, Stockholm, Sweden\\
$^{2}$Department of Physics and Astronomy, Uppsala University, Box 515, SE-751 20 Uppsala, Sweden\\
}

\date{Accepted 2019 April 29.}

\pubyear{2019}

\begin{document}
\label{firstpage}
\pagerange{\pageref{firstpage}--\pageref{lastpage}}
\maketitle

\begin{abstract}

We have used the capability of the Multi-Unit Spectroscopic Explorer (MUSE) instrument to explore the impact of stellar feedback at large scales in Haro 11, a galaxy under extreme starburst condition and one of the first galaxies where Lyman continuum (LyC) has been detected. 
Using \Ha , \OIII\ and \OI\ emission lines from deep MUSE observations, we have constructed a sequence of velocity-dependent maps of the \Ha\ emission, the state of the ionized gas and a tracer of fast shocks. These allowed us to investigate the ionization structure of the galaxy in 50 \kms\ bins over a velocity range of -400 to 350 \kms.
The ionized gas in Haro 11 is assembled by a rich arrangement of structures, such
as superbubbles, filaments, arcs, and galactic ionized channels, whose appearances change
drastically with velocity. The central star-forming knots and the star-forming dusty arm are the
main engines that power the strong mechanical feedback in this galaxy, although with different
impact on the ionization structure.
Haro 11 appears to leak LyC radiation in many directions.
We found evidence of a kpc-scale fragmented superbubble that may have cleared galactic-scale channels in the ISM. Additionally, the Southwestern hemisphere is highly ionized in all velocities, hinting at a density bound scenario. A compact kpc-scale structure of lowly ionized gas coincides with the diffuse \La\ emission and the presence of fast shocks. 
Finally, we find evidence that a significant fraction of the ionized gas mass may escape the gravitational potential of the galaxy. 

\end{abstract}




\section{Introduction}

Blue compact galaxies (BCGs) play a fundamental role in our understanding of the evolution of galaxies. They are compact, mostly metal poor galaxies, that are undergoing an extraordinary episode of star formation, characteristics that are similar to the primeval high redshift galaxies
\citep{Fanelli1988,Oestlin2001,Thuan2005,Wu2006,Thuan2008,Kunth2000}. 

Their current burst of star formation seems to be triggered mainly by infalling H I clouds or by gas compression in merger systems \citep{Oestlin2001}. The extreme starburst condition favours the formation of massive star clusters, or even super star clusters (SSCs, M$_{cl} > 10^5$ \Mo ) \citep{Oestlin2003,Adamo2011,Bik2018}, each of them containing a large number of massive stars, whose stellar feedback has strong implication in the subsequent evolution of the galaxy.

In the first $\sim$ 4 Myr of a star cluster evolution, OB-type stars release large amounts of ionizing photons into the ambient medium before they explode as supernovae. This is the fraction of time in the star cluster evolution, where the ionization and radiation feedback are at their maximum. At the same time, their stellar winds inject kinetic energy and momentum into the surrounding medium.
The mechanical energy is maintained afterwards till $\sim$ 40 Myr by supernova explosions that additionally deliver newly synthesized elements to the environment \citep{Leitherer1999}.

Stellar feedback, and especially mechanical feedback, produces shock waves that may trigger additional star formation, or suppress it by lowering the neutral gas content from the active zone of star formation \citep{Dale2005, Hopkins2014, Krumholz2014}, and thus, regulating the accelerated consumption of the cold gas. Moreover, it induces turbulence in the interstellar medium (ISM) supporting the formation of a porous medium that increases the mean free path of the ionizing photons, enabling them to penetrate further into the halo \citep{Silk1997,Zurita2002,Bagetakos2011}.

The effect of stellar feedback is not well understood in low-metallicity galaxies. Numerical works show that in low-mass galaxies, stellar feedback can be powerful enough to drastically modify the structure of the ISM \citep{Ceverino2009,Hopkins2012,Agertz2013,Keller2015}. The injected kinetic energy can develop a violent
ISM, supporting the formation of large-scale bubbles, filaments, arcs, loops, rings, and cavities (see \citealt{Tenorio-Tagle1988,Hunter1993,Hopkins2012}).
These structures are evident in several galaxies such as M82 \citep{Lynds1963,Bland1988}, NGC 1705 \citep{Meurer1992}, NGC 3079 \citep{Cecil2001} and ESO 338-IG04 \citep{Bik2015,Bik2018}. Moreover, \citet{Marlowe1995} and \citet{Martin1998} studied the kinematics and morphology in a small sample of starburst dwarf galaxies and found in the vast majority, footprints of expanding superbubbles, traced by fragmented super-
shells and filaments.
\citet{Hunter1993} concluded that, at least one of these structures caused by feedback, may be seen in the ISM
of a large fraction of luminous low-mass galaxies.

In the framework of galactic winds, these bubbles and superbubbles may drive large-scale outflows when they fragment \citep{Chevalier1985,Tenorio-Tagle2006,Heckman2017,Fielding2018}.
Supernova-driven galactic winds can accelerate ambient gas at velocities much greater than their escape velocities, resulting in a loss of metal-enriched gas to the intergalactic medium. This process could explain
the deficit of metals in dwarf galaxies \citep{Andrews2013,Sanchez2013}.  
Although a large number of galaxies display fast outflows, only few galaxies develop galactic outflows with
velocities greater than their escape velocities. Most of them have been inferred from absorption lines tracing the neutral and ionized gas phase \citep{Chisholm2015,Heckman2015}.

Galactic-scale outflows also play a fundamental role in the escape of Lyman continuum (LyC) radiation by creating galaxy-scale holes in the ISM favouring the escape of LyC photons \citep{Fujita2003}.
LyC emission is rarely detected in galaxies, mainly because the neutral hydrogen column density along the line of sight to their production places is high enough to prevent LyC photons to escape. 
To date, only a small numbers of galaxies (16) have been found to leak LyC radiation \citep{Bergvall2006,Borthakur2014,Vanzella2016,Barros2016,Izotov2016a,Izotov2016,Leitherer2016,Puschnig2017,Vanzella2018,Izotov2018,Izotov2018a}.
Although the mechanism favouring the escape of LyC radiation is still unknown, galactic holes possibly cleared by outflows may have preference from a density-bound medium, due to the relative high neutral column density measured in the ISM of those galaxies \citep{Chisholm2015,Vanzella2018}

The galaxy studied in this paper is Haro 11, a well-known starburst luminous blue compact galaxy. Haro 11 is a \La\ emitter \citep{Hayes2007} and one of few galaxies where Lyman continuum has been detected 
\citep{Bergvall2006,Leitet2011}. 
In contrast to most BCGs, its ionized gas mass is larger than its neutral gas content \citep{MacHattie2014,pardy2016}.
Morphologically, it is a merger system whose appearance and kinematics resemble the Antennae galaxy \citep{Oestlin2015}.

Star formation happens mostly in three knots A, B and C and a dusty arm at the centre of the galaxy (see Fig. \ref{fig:Ha_HST_center}) \citep{Kunth2003}.
The star cluster population in Haro 11 is very young and massive. 
\citet{Adamo2010} identified around 200 clusters with masses ranging from 10$^4$ to $\sim$10$^7$ \Mo\ . Most of them (90\%) were formed in the current starburst that started 40 Myr ago \citep{Adamo2010,Oestlin2001}.
In half of them, supernova explosions may have recently started, since they were formed at the peak of the cluster formation, around 3.5 Myr ago.
Thus, we are capturing Haro 11 at a time when the radiative and mechanical energy released by its massive stellar population is at its maximum.
Beside the stellar components, \citet{Prestwich2015} detected hard X-ray emission in Knot B that hints at an intermediate black hole in low accretion mode. 

The intense starburst condition and merger dynamics have strong impact in the kinematics of this galaxy. Several kinematic components have been reported at all wavelengths. 
\citet{Grimes2007} found two main outflows at -80 \kms\ (FWHM $\sim$300 \kms ) and -280 \kms\ in the warm UV gas, that are associated with outflowing winds.  
\citet{Rivera-Thorsen2017} found evidence for a clumpy neutral ISM.
The \La\ morphology analyzed by \citet{Hayes2007} hints to the presence of a bipolar outflow at the base of Knot C. Recently \citet{pardy2016} found that the neutral 21 cm HI gas is moving at +56 \kms\ (FWHM  of 77 \kms ). In the optical range, the warm ionized gas was found to have a multi-component nature by \citet{Oestlin2015}, with velocities ranging from -130 to 130 \kms .
These studies give insight into the complex kinematics of this merger system.

In this paper we examine the impact of stellar feedback in Haro 11 by means of analysing the velocity resolved 
\Ha\ emission, the level of ionization in the gas and possible presence of fast of shocks. In our analysis, we use the capability of MUSE in combining the spectral and spatial high quality information at the same time, allowing us to spatially resolve the ionized gas structures in the galaxy. 

This paper is organised as follow: the observations, data reduction and methods are presented in Section 2. In Section 3 we describe the most prominent structures found in the ionized gas. In section 4 we discuss the ionized gas structure and analyse the mechanism that might have facilitate the escape of LyC photons in Haro 11. Additionally, we estimate the ionized gas mass that may escape the gravitational potential in this galaxy.  And last, we present a summary and conclusion in section 5.

In this paper, we use the redshift of Haro 11 of z$_{Haro11}$=0.020598 from NASA/IPAC Extragalactic Database, a Hubble constant of H$_0$=73 \kms\ Mpc$^{-1}$ and matter density $\Omega_0$= 0.27) resulting in a luminosity distance of 82 Mpc.

\section{Observations and data reduction}

\subsection{Observations}

\begin{figure}  
    \centering
    \includegraphics[width=5cm]{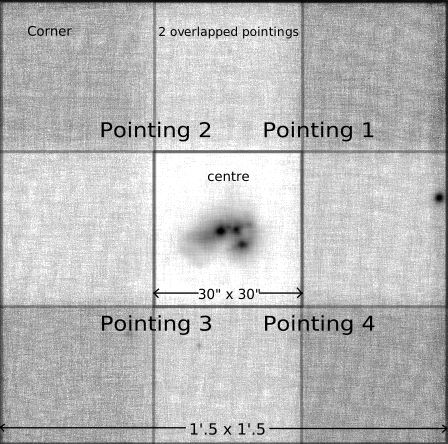}
    \caption{MUSE noise map of a random spaxel showing the mosaic configuration of the observed field.
    }
    \label{fig:mosaic}
\end{figure}

Haro 11 was observed with the Multi-Unit Spectroscopic Explorer \citep[MUSE,][]{Bacon2010} at the Very Large Telescope on Paranal, Chile between December 2014 and August 2016. To cover the extended ionized halo of the galaxy, a 2x2 mosaic (see Fig. \ref{fig:mosaic}) centred on Haro 11 (RA=00$^h$36$^m$52$^s$ DEC=-33$^\circ$33\arcmin17\arcsec) was performed. Adjacent borders overlap by 30\arcsec. Thus, the 1\arcmin.5  x 1\arcmin.5 full field observed corresponds to 39.1 x 39.1 kpc$^2$ at the redshift of Haro 11. Each pointing was observed in two observing blocks 
with four frames of 700s and position angles rotated by 90$^\circ$ to correct for detector systematics. The design of the mosaic resulted in a final cube with three areas of different integration time
While the corners have an integration time of one pointing i.e. 1h 33min 20s (4x700s), the two overlapped regions have twice that integration time, 3h 6min 40s (8x700s) and the central 30\arcsec x 30\arcsec region has the longest integration time of 6h 13min 20s (16x700s). 

The observations were carried out under good atmospheric conditions (photometric or clear) with seeing measurements varying between 0.6 and 0.9 arcsec. The final image quality (FWHM of a Gaussian fit) measured on the white-light image from the reconstructed cube is about 0\arcsec.8. 
The spectrophotometric standard star Feige 110 was observed in the same nights and is used for the spectrophotometric calibration of the data.

\subsection{Data reduction}
 
We apply the standard reduction procedure using the MUSE pipeline version 1.2 \citep{Weilbacher2012} with a minor change for the sky subtraction process.
To subtract the sky from each science exposure, a sky spectrum can be either provided as input file, in case sky
exposures were taken, or is created from the darkest areas in the science exposures. As no sky exposures were obtained during the observations of Haro 11, we calculate the sky from the science exposures by generating a sky mask. 
Given the large extent of the galaxy halo, usually reaching the edges of the field of view, and the
small regions at the corners left for the sky, the extracted sky mask always encloses the weakest areas of the ionized halo, resulting in over subtraction of it. To avoid this over subtraction, we perform a two-step process:

First, we create a sky mask for each exposure selecting the 20\% darkest pixels in the first run of the recipe \texttt{muse\_scipost}. Then, we customise the sky mask manually to remove areas of the H$\alpha$ and [OIII]$\lambda$5007 emission, where the halo has the largest size. Last, the \texttt{muse\_scipost} recipe is run again with the modified sky mask as input file, resulting in a correct sky subtraction.

Finally, we create the final cube by combining all the reduced exposures. The extended configuration of MUSE covers the wavelength from 4600 {\AA} to 9350 {\AA}  with a spatial sampling of 0\arcsec.2 x 0\arcsec.2 per spaxel and a spectral sampling of 1.25 {\AA}.

\subsection{Correction for stellar absorption}

Galaxies with young- to intermediate-age stellar populations show broad stellar absorption, which are underlying the emission lines in the Balmer series. This effect originates in the atmosphere of primarily type A stars and increases towards the higher energy levels of the Balmer series \citep{GonzalezDelgadoI1999,GonzalezDelgadoII1999}.

In Haro 11, this stellar absorption is particularly strong towards the southeast (SE) - east (E) of knot C, where the evolved progenitor is located. In the MUSE spectrum, we can see that the absorption is especially noticeable in the wings of the line.
In order to have correct emission line strengths for accurate line ratios, the underlying stellar absorption affecting the Balmer lines needs to be removed.

We fitted the spectra of each pixel 
using the python package of the penalized pixel-fitting method (\texttt{pPXF}) by 
\citet{Cappellari2017} with the \texttt{MILES} stellar spectra library \citep{Sanchez-Blazquez2006, Vazdekis2010}
 in the regions where the stellar continuum has a signal-to-noise ratio (S/N) above 20. 
The strongest emission lines were masked to improve the fit on the absorption features. For those pixel, where the fitting resulted in a good fit ($\chi^2$ < 5), the best fit was subtracted from the observed spectrum. This procedure removed all the absorption line contamination leaving a emission line spectrum only.

\subsection{Velocity resolved emission line maps}

\begin{table}
    \centering
    \begin{tabular}{lcccc}
    \hline
    Emission     &  $\lambda_{Haro\_11}$ & Spaxels in & Spectral  & Spatial  \\
    line         &               &    50 kms$^{-1}$   & resolving & resolution \\
    				 &  [\AA]    & [\AA]  & power & $[\arcsec]$ \\
    \hline
    H$\beta$                    & 4961.4     & 0.83 & 3.5  & 0.94 \\
    $[$OIII$]\lambda$5007         & 5109.9   & 0.85 & 3.4  & 0.92 \\
    $[$OI$]\lambda$6300           & 6429.8   & 1.07 & 2.3  & 0.85 \\ 
    H$\alpha$                    & 6697.9    & 1.12 & 2.1  & 0.84 \\
    \hline
    \end{tabular}
    \caption{Spectral and spatial resolution of the emission lines 
    after resampling.}
    \label{tab:resampling}    
\end{table}

The full width at half-maximum (FWHM) of the observed line spread function (LSF) of MUSE is wavelength dependent and varies between 2.9 \AA\ for \Hb\ and 2.5 \AA\ for \Ha . This effect becomes very important when analysing line ratio diagnostics in velocity bins and needs to be corrected. To do so, we convolved the spectrum of the \Ha\ line with a Gaussian kernel sampled from the spectral resolution difference of the \Ha\ and \OIII\ lines. These \Ha\ corrected spectra was used in the \OIIIHa\ diagnostic. The \Ha\ and \OI\ lines are close in wavelength and have similar spectral resolution, therefore it was not necessary to correct the \Ha\ for the \OIHa\ diagnostic.

From the corrected data cube, we extracted the spectra of the lines \Ha, \Hb, \OIII\ and \OI\ (hereafter [OIII] and [OI]) and re-sampled them using the python package \texttt{SpectRes} \citep{Carnall2017} to 50 \kms\ per resampled spaxel. The resampled spaxels have a width of 0.8\AA\ for \Hb\ and 1.1 \AA\ for the \Ha\ line (see Tab. \ref{tab:resampling}), smaller than the original spectral sampling of MUSE (1.25 \AA\ per spaxel). 
Comparing the spectral resolving power of the MUSE instrument to the velocity resolution of the interpolated spaxels, shows that due to the LSF of the instrument a spectrally unresolved feature will be spread out over 2 (\Ha )  to 4 (\Hb ) velocity bins.
Table \ref{tab:resampling} summarizes the values for each of the extracted emission lines.
Finally, we extract line maps and their associate uncertainties from continuum-subtracted line fluxes per velocity bins for all four lines.
We did not correct the line maps for extinction. The halo shows E(B-V) values between 0.02 and 0.1, indicating that there is little or no extinction. The highest E(B-V) values were measured mostly at redshifted velocities in a small area around knot C and B and the dusty arm. 
We have verified that correcting our maps will not change our analysis that is dedicated in describing the - mostly in kpc-scale - ionized gas structures in Haro 11. However, it will add a value of 0.25 to the \OIIIHa\ ratios in the regions with the highest extinction. These corrections do not change the condition of the ionized gas in these regions.

Since the seeing is wavelength dependent, we measure the spatial resolution of each line by performing a Gaussian fit and measuring the full width half maximum (FWHM) (see Tab. \ref{tab:resampling} col. 5) in a bright distant star in the west part of the field of view. 

To unify the spatial resolution of the [OIII] and H$\alpha$ for the ionization diagnostic, we convolved the maps of the emission line with higher spatial resolution, the H$\alpha$ line, with a Gaussian kernel sampled from the resolution difference of both lines. The spatial resolution of the [OI] maps are close to the spatial resolution of the H$\alpha$ maps, therefore we do not convolve the [OI] line for the [OI]/\Ha\ line ratio.
Finally, we created velocity sliced ionization maps from the [OIII]/H$\alpha$ line ratios and the maps tracing shocks from the [OI]/H$\alpha$ line ratios.

To keep a minimal signal to noise in the data, we masked the emission line regions in all maps,
by adaptively Voronoi binning each map with a signal to noise of 4 and a maximal area of 1.\arcsec 8 \citep{Cappellari2003,Diehl2006}. Cells with signal-to-noise ratio greater or equal are removed, leaving only the sky regions unmasked. This mask was then applied to the unbinned maps, removing the background noise and low S/N regions.

\section{Condition of the ionized gas at different velocities}

\begin{figure*}
    \centering
    \includegraphics[width=2\columnwidth]{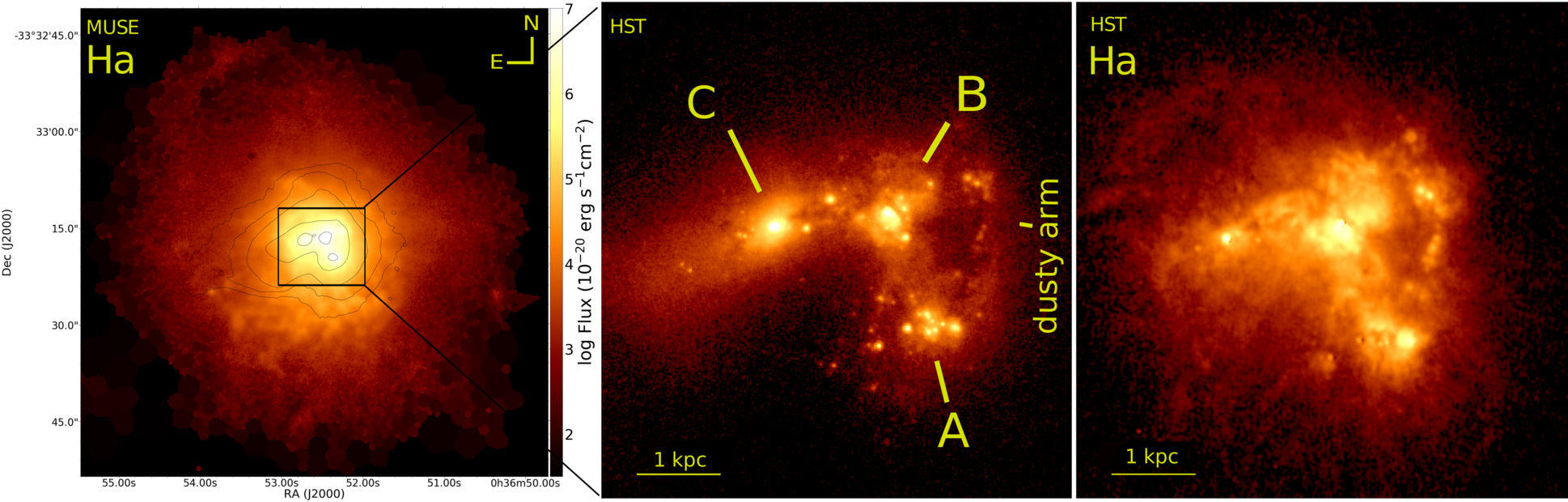}
    \caption{Haro 11 as seen from MUSE and HST data. The left image shows the MUSE H$\alpha$ map displaying the full extent of the ionized halo. The middle and right images show the central 5x5 kpc of the galaxy that were taken with the HST with the medium R and \Ha\ filters respectively. The first and third images show clearly the difference of the \Ha\ image taken by the HST and constructed with MUSE data. The central star forming knots A, B and C are labelled following the nomenclature of \citet{Kunth2003} 
    }
    \label{fig:Ha_HST_center}
\end{figure*}

Fig. \ref{fig:Ha_HST_center} shows Haro 11, seen by MUSE and the Hubble Space Telescope (HST).  
The left image displays the integrated \Ha\ map (Voronoi binned to a S/N$\geq$5 in \Ha ) from the MUSE cube. Haro 11 exhibits a huge ionized halo, that extends over 30 kpc in diameter down to a sensitivity level of 3.75 \x 10$^{-19}$ erg s$^{-1}$ cm$^{-2}$ arcsec$^{-2}$. In addition to the overall spherical distribution of the ionized gas, there are some faint noticeable features:
for instance, the  arc in the north-east and the network of filaments and clumps in the Southern hemisphere.
Later in this section we will show that these and other features stand out clearly in the velocity resolved \Ha\ maps.

The images in the middle and right panels in Fig. \ref{fig:Ha_HST_center} were taken with the HST and show the optical continuum (red part, F763M medium filter) and the continuum subtracted \Ha\ emission, respectively.
They show the arrangement of stars (central image) and warm gas (right image) in the central 5 $\times$ 5 kpc of the galaxy.
The central knots A, B and C and a the dusty arm (dubbed the 'ear' in \citealt{Oestlin2015}) are the most active zones of star formation.
The dust, which is not well traced in the images, is located mainly in the dusty arm and in dust lanes crossing knots B and C.

The contrast of the \Ha\ map taken with the HST and constructed from MUSE data is dramatic. HST provided a high spatial resolution image that shows the central part of Haro 11 in great detail, while our MUSE image has higher sensitivity in a larger field of view.

To investigate the structure of the ionized gas in details, we further use ionization maps and a tracer of fast shocks. Ionization maps are constructed from nebular emission line ratios of two ions with different ionization potential. They trace the level of ionization in a nebula and the strength of the UV radiation field. We use the ratio of the two most intense optical emission lines, [OIII]$\lambda$5007/\Ha , as they trace the halo to the furthest. Since the ionization potential of hydrogen is significantly lower (13.6 eV) than that of the double
ionized oxygen (35 eV), a rise in the [OIII]$\lambda$5007/\Ha\ ratio implies therefore, a rise in the number of energetic photons capable of double ionize oxygen.

The forbidden \OI\ line is widely used to trace the radiation of the ISM heated by shocks. Neutral oxygen atoms are
found in a neutral and weakly ionized gas. Thermal electrons of low energy ($\sim$1.9 eV) are able to excite these atoms through collisions and produce the observed [OI]$\lambda$6300 line.
Shocks can increase the thermal energy of the free electrons in proportion to the shock velocity, and thus enhancing the number of collisional excitations by a single electron.
Consequently, higher [OI]$\lambda$6300/\Ha\ ratios are expected to arise in the ISM where intermediate to strong shocks are present \citep{Veilleux1987}.

Fig \ref{fig:HaIS} presents the velocity sliced ionized gas structures in Haro 11. 
The columns from left to right show: the ionized gas architecture(traced by \Ha\ emission), the degree of ionization of the ISM ([OIII]$\lambda$5007/\Ha ) and the regions where intermediate to fast shocks are present ([OI]$\lambda$6300/\Ha ).
The I-band continuum emission from the MUSE data is overplotted in contours. These are shown to provide a spatial reference in the maps. 
Each gas diagnostic is extracted in velocity bins of 50 \kms\ (rows in fig. \ref{fig:HaIS}) starting from -400 to 350 \kms . These velocities correspond to the limits of the \Ha\ wings uncontaminated by the broad [NII] lines.  
Since the [OIII]$\lambda$5007 is uncontaminated by strong lines, we extract additional maps of this line to velocities up to 1000 \kms . These are referred in Section 4.  

Below we describe each of the diagnostics from high blueshifted velocities (-400 \kms) towards high redshifted velocities (350 \kms ) in further detail. The number of features is too large for each one to be individually described, but we list some of the most prominent and give a general description of the structure seen at
different velocities. 
A summary of the features seen in the \Ha\ and ionization maps are listed in Table \ref{tab:StructuresHa_Map}.

\subsection{Velocity sliced \Ha\ maps}

\begin{table*}
    \centering
    \begin{tabular}{lcccl}
    \toprule
    Structures                         & length / (*) radius               & \multicolumn{2}{c}{Velocity}& Orientation              \\
    in the 	\Ha\ maps			  & (**) Dist. to knot-C 				& \multicolumn{2}{c}{Range}  &                        \\
    								  & [kpc]	                   & \multicolumn{2}{c}{\kms\ }     &         \\                 
    \hline                 
    Tidal tail 1,2 and 3             & 10, 5.6, 18.4                          & -100           & 100 & NE, E, from SE to E            \\
    Filaments                         & 11                          & -350           & $\sim$50 & Northern hemisphere                  \\
    Luminous arc (Supershell ?)          & 3.3 (*)                         & -400          & $\sim$ 50   & Centred in the central knots.   \\
    6 compact clumps in the halo & 18.8,  17.3, 15.6 (**)          & 0            & 50  & End of the tidal tail  \\
        							  & 7 (**)                          & 50            & 150  & S -- E.  In arc 2 or tidal tail  \\
        							  & 5.4 , 4.7 (**)                  & -400          & -250 & S and NE  \\      
    blob                            & 5.6                         & 200              & 350 & NE                   \\                                      
                                      &                                 &               &      &           \\
    \bottomrule     
    Structures                         & level               & \multicolumn{2}{c}{Velocity}&               \\
    in the 				              & of				& \multicolumn{2}{c}{Range}  &     Direction                   \\
    ionization maps 				  & ionization	                   & \multicolumn{2}{c}{\kms\ }     &         \\                 
    \hline                 
     Southwestern hemisphere & very high                          & -350           & 100      & Southern hemisphere            \\         
     Western hemisphere & very high                          & 100           & 350      & Western hemisphere            \\       
    Second superbubble centred in knot C  & low                         & -300           &  & Centred in knot C. Probably reappears at $\geq$ 300 \kms      \\
        Superbubble - Supershell          & high and low                 & 300          & 350   & Centred in the star forming knots (indirect evidence).   \\
    Filaments                         & low                            & -50          & 50 & Southern hemisphere                 \\
    2 broad channels              & highest                               &-250           &      & S         \\    
    3 channels (conical shape) & high              & -100          & 100   & N and E (2 of them)  \\ 
     							&               &  100          & 250   & SW   \\
    2 narrow channels & high              & -250          & -150   & NW  \\                
    Large structure            & low              & -200          & 200  & Moves with velocity from E to N.    \\
    \bottomrule                                    
    \end{tabular}
    \caption{Summary of the prominent features found in the \Ha\ and \OIIIHa\ maps. In this paper, low ionized regions have \OIIIHa$\leq$0.7 and the high ionized have \OIIIHa$\geq$1.8 }
    \label{tab:StructuresHa_Map}      
\end{table*}

The \Ha\ emission shows plenty of structures, many of which only become visible when we slice the galaxy in velocity space. 
The most prominent features by their \Ha\ intensity are the three starburst knots: A, B and C. 
They are visible in all velocity bins, but each with their own characteristics. Knot C can be seen as a compact
source with a broad velocity width. Knot B is a complex structure whose morphology changes with velocity, while knot A becomes more prominent at positive velocities.

From -400 to $\sim$50 \kms , 
an \textit{arc} develops in the northeast (NE) at a radius of $r\sim$3.3 kpc and is best visible at -300 \kms\ where it subtends an angle of 180 degrees. This arc is also visible in the HST \Ha\ map seen in the third panel of the Fig. \ref{fig:Ha_HST_center}.

In the same velocity range, a group of winding \textit{filaments} develop somewhat radially in the south-southwest (S-SW) at $r>3$kpc and bend slightly towards the east (E) and west (W) in a symmetrical fashion. At redshifted velocities, the filamentary structure is not clearly disentangled from new (oncoming) structures.
These filaments extend to the edges of the observed emission and cover the entire S-SW halo, while the northern halo is nearly free of structures and its intensity drops considerably with radius, rapidly becoming diffuse and faint.

From -400 to -250 \kms , two compact \textit{clumps}  become visible in the halo and may be locally ionized by their young stellar components. We detected these sources in the blue (HST filter F336W) and weakly in the red continuum (HST filter F763M).

At velocities from -100  to 100 \kms , but best visible at 50 \kms , three \textit{tidal tails} (two displaying an arc-shape) develop at the east, northeast and east part of the galaxy respectively. 
From 50 to 250 \kms\ a structure that seems to belong to the \textit{tidal tail 2} or the \textit{tidal tail 3} becomes pronounced. It whirls from the SE to the S of the galaxy. Within this structure, from 50 to 150 \kms\ a bright and compact \textit{clump} becomes visible.

At 0 and 50 \kms\ three faint \textit{clumps} appear at the farthest side of the tidal tail 3 at distances > 15 kpc.
The closest one is weakly traced in the \OIII\ line, while the remaining two are not observed. These clumps might be locally ionized by their stellar components or in case of the furthermost, they could be ionized by photons originated from the central clusters of the galaxy.
At high redshifted velocities (v$\geq$200 \kms ) a \textit{blob} develops in the NE.

\subsection{Velocity sliced ionization maps}

The halo is highly ionized over all velocities, although the areas with the highest level of ionization change with velocity. 
In the inner part (r<10 kpc) of the galaxy, there is plenty of kpc-scale lowly ionized structures that contrast over the bright highly ionized halo. Some features resemble the structures seen in the \Ha\ maps, while new ones arises. 

From -350 to 100 \kms\ the southwestern hemisphere is highly ionized and is particularly enhanced at -250 \kms\ in two  \textit{highly ionized channels}, reaching [OIII]/\Ha\ ratios up to 3.5. In the \Ha\ maps, this part of the halo is entirely populated by the filamentary structure. Moreover, the areas between the filaments are in general
more enhanced and suggest regions of low-density gas.

At v $=$ -300 \kms , a system of circumferentially oriented lowly ionized \textit{arcs} define what seems to be a slightly fragmented \textit{shell} of $\sim$1.7 kpc of radius that is centred on the massive star forming knot C.
Its interior is highly ionized, suggestive of a low-density hot gas.
When comparing the position of the lowly ionized arcs with the \textit{arc} seen in the \Ha\ map, the latest seems to be located slightly more outside. 
At v$\geq$ 300 \kms\ part of this lowly ionized arc system reappears and it probably shows the same structure in its farthest side. We note also a narrow lowly ionized pillar crossing knot C, that could trace outflows of neutral and low-ionized gas.

From -250 to -150 \kms , two narrow and slightly bent highly ionized channels stand out over a lowly ionized structure (see also Fig. \ref{fig:details} b). These channels seem to be tracing the footpath of \textit{outflows} 
which are carrying the hot high ionized gas from the knot C superbubble cavity to the halo.

From -200 to 200 \kms , the lowly ionized gas is mainly condensed in a kpc-scale compact structure that turn with velocities from the SE to the N and end at v > 200 \kms\ in the \textit{blob} identified in the \Ha\ maps. 
From 100 to 200 \kms , but best visible in Fig. \ref{fig:details} a, his lowly ionized structure becomes multistructured.
Its morphology resembles somehow a fishing hook, although this effect could be caused by projection effect of lowly ionized gas clouds along the line of sight. 

From -100 to 100 \kms , three \textit{highly ionized channels} become visible towards the N, W and the S of the halo. In Fig. \ref{fig:details} a\ ), the northern channel can be traced to the place where it originates, at the
basis of knot B. The western and southern channels however might be created by photons produced in knot A and B. Additionally, there are few smaller highly ionized channels (see Fig. \ref{fig:details} a\ )at the top of the
lowly ionized structure that seem to be created by photons leaking from knot C.

From -50 to 50 \kms\ at central velocities, the  \textit{lowly ionized structure} predominate over a highly ionized halo. In this velocity range, the most prominent filaments become lowly ionized likely due to an increase in their gas density, rising their neutral and lowly ionized gas content.

From 100 to 350 \kms , the western hemisphere is highly ionized and especially enhanced at 250 \kms towards the SW, where one of the highly ionized channel elongates.

\begin{figure*}
    \centering
    \begin{subfigure}[b]{1\textwidth}
    \centering
    \includegraphics[width=0.8\columnwidth]{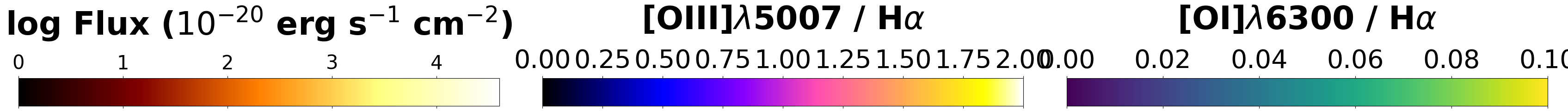}
    \label{fig:CB}
    \end{subfigure}
    \begin{subfigure}[b]{1\textwidth}
    \centering
   \includegraphics[width=0.8\columnwidth]{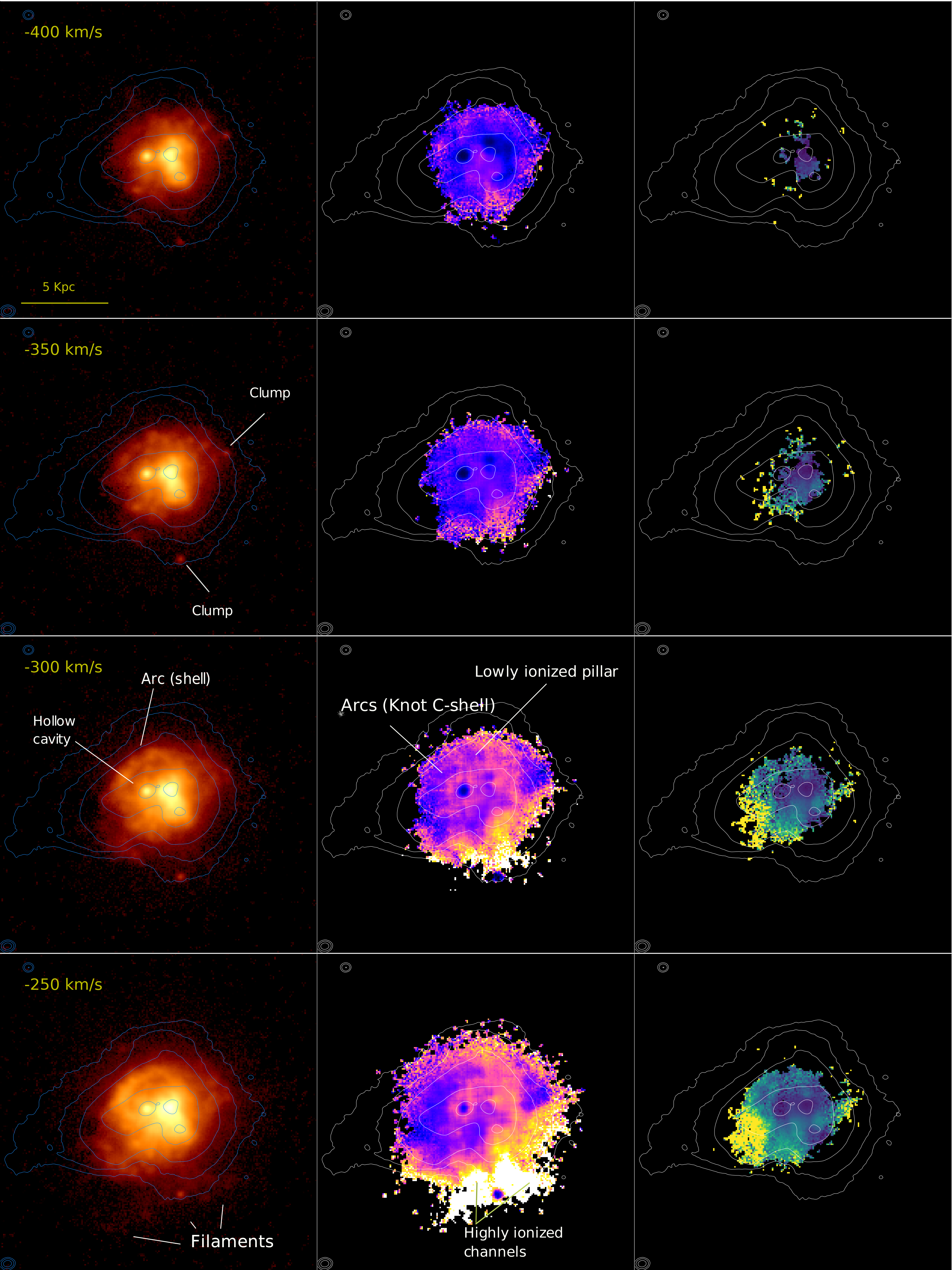}
    \end{subfigure}
    \caption{Velocity sliced \Ha\ maps (left) of Haro 11. The overlaid contours show the MUSE I-Band stellar continuum in blue. The isophotal level of the faintest contours is 7.5 $\times$ 10$^{-19}$ erg s$^{-1}$ cm$^{-2}$ arcsec$^{-2}$ \AA$^{-1}$. The middle column shows the [OIII]/\Ha\ ratio tracing the ionization of the gas, while the right column shows [OI]/\Ha\ highlighting the shocked gas. The colourbar of the maps in each emission line diagnostic is fixed to the same scale and limit parameters. For the \Ha\ maps, the lower surface brightness limit of 2.5 $\times$ 10$^{-19}$ erg s$^{-1}$ cm$^{-2}$ arcsec$^{-2}$ at 50 \kms . For the [OIII]/\Ha\ and [OI]/\Ha\ maps, the limits range from 0 to 2 and to 0.1 respectively. The maps in the velocity range from -100 to 100 \kms\ are displayed in the full field of view. The yellow boxes in the \Ha\ maps show the zoom in region displayed on the the maps at higher blue- and redshift velocities. A small area towards the east covering the clumps in the 50 \kms\ \Ha\ map is shown with an intensity one hundred times higher to highlight the presence of these clumps. } 
    \label{fig:HaIS}
\end{figure*}

\begin{figure*}\ContinuedFloat
    \centering
    \includegraphics[width=2\columnwidth]{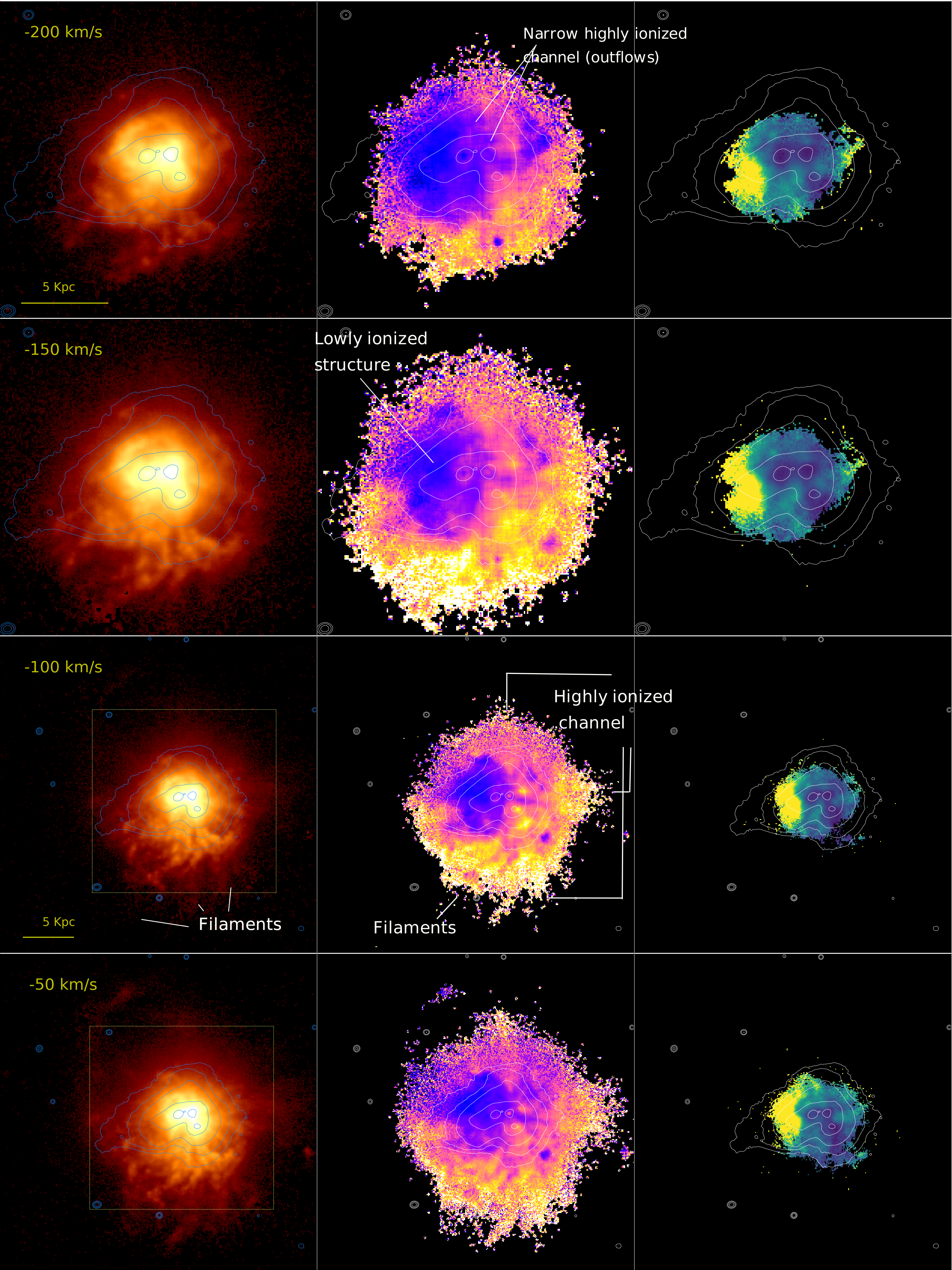}
    \caption{Ibidem. Gas diagnostics maps for velocities from -200 to -50 \kms\ .     }
\end{figure*}

\begin{figure*}\ContinuedFloat
    \centering
    \includegraphics[width=2\columnwidth]{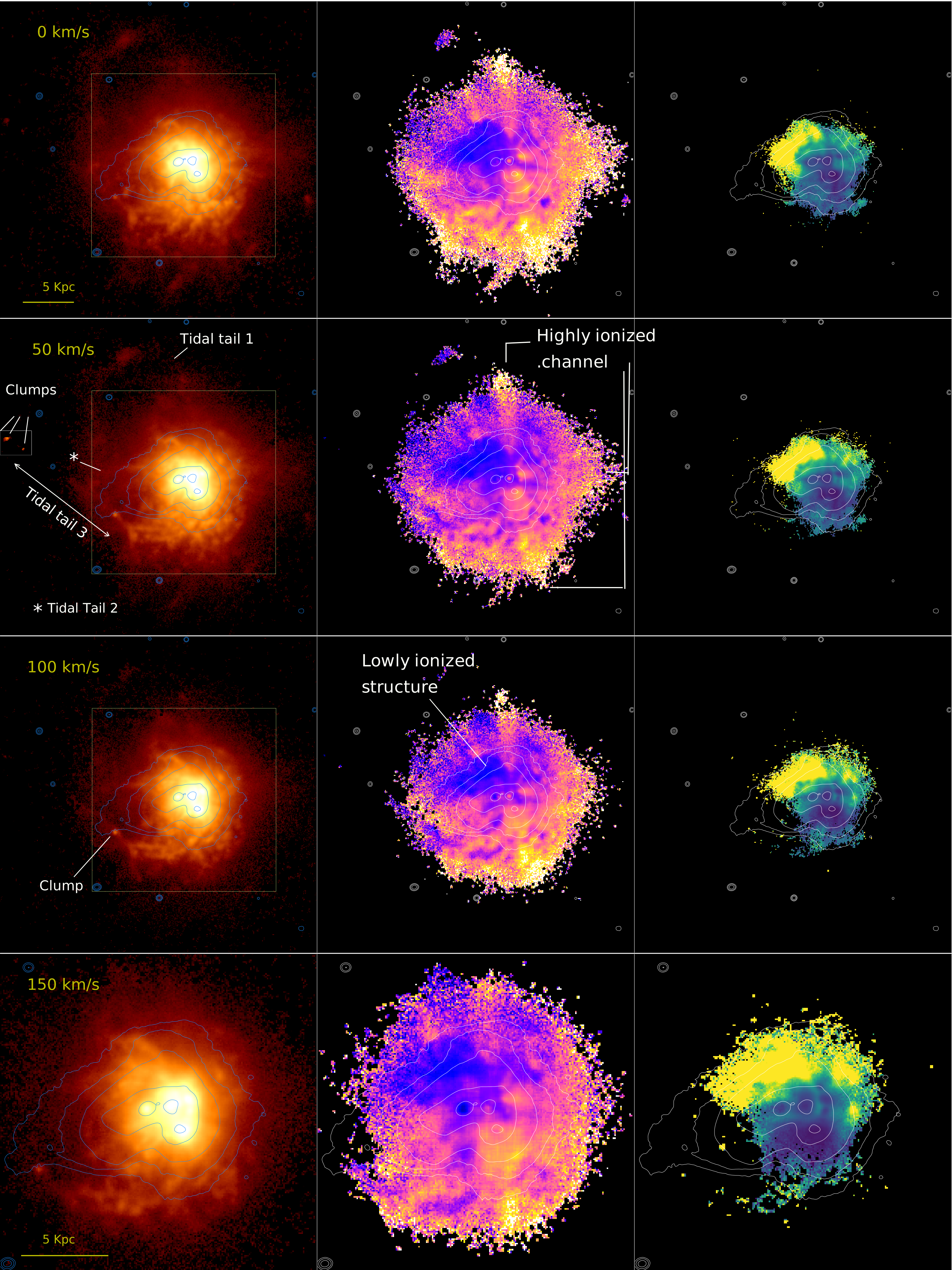}
    \caption{Ibidem. Gas diagnostics maps for velocities from 0 to 150 \kms\ .     }
\end{figure*}

\begin{figure*}\ContinuedFloat
    \centering
    \includegraphics[width=2\columnwidth]{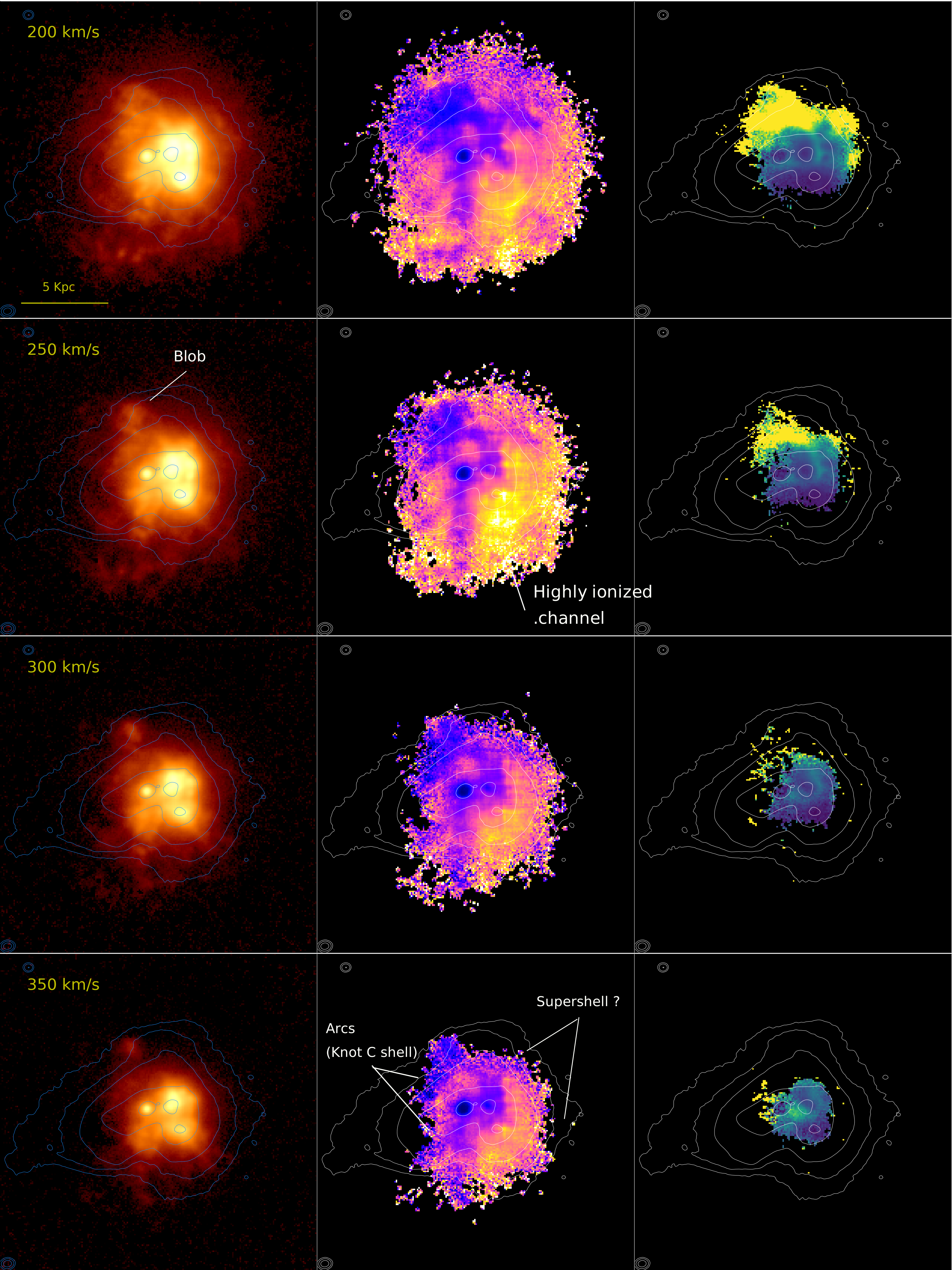}
    \caption{Ibidem. Gas diagnostics maps for velocities from 200 to 350 \kms\ .}
\end{figure*}

\begin{figure*}
    \centering
    \includegraphics[width=2\columnwidth]{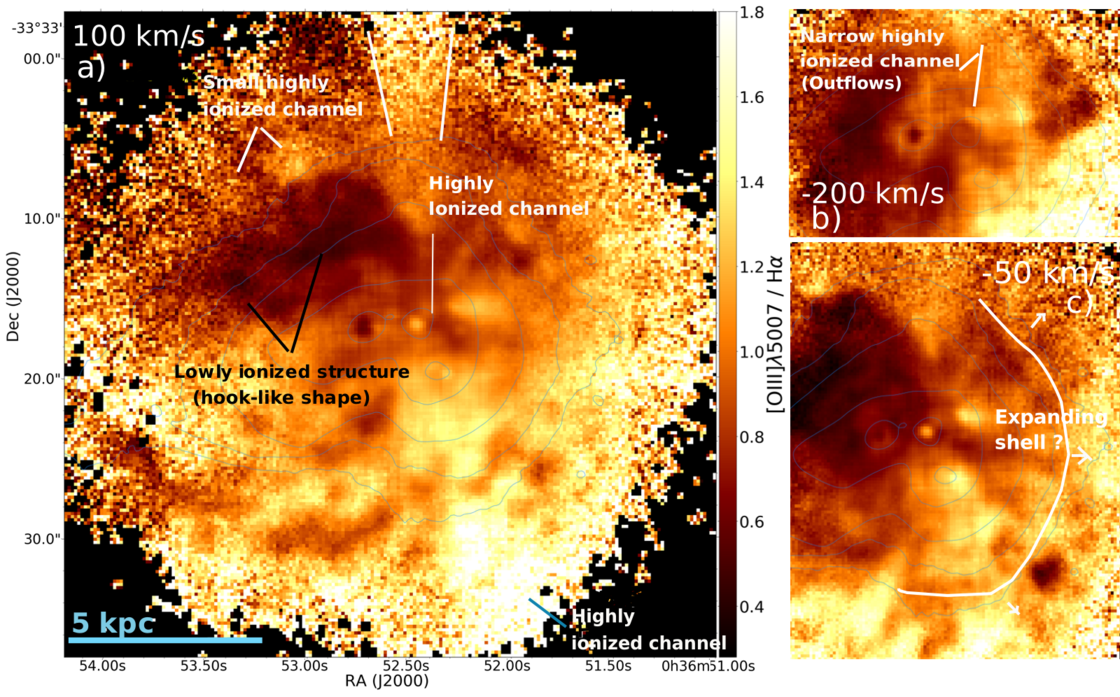}
    \caption{Three ionization maps at velocities of 100, -200 and -50 \kms\ with higher contrast in the colour map (scale range from 0.3 to 1.8) highlighting the features of feedback origin. 
    The left (a) and top right (b) panels show highly ionized channels with sizes ranging from 1 to 10 kpc. The narrow highly ionized channels of (b) are most likely created by outflowing hot gas from knot C, while the large conical channels of (a) seem to be created by huge amount of ionizing photons released by the central knots. Moreover, panel (a) shows the lowly ionized structure with a hook-like shape in this velocity bin. The bottom right panel (c) shows circumferentially oriented arcs, resembling an expanding shell.}
    \label{fig:details}
\end{figure*}

\subsection{Fast shocks}

The \OI\ faint line traces only the high S/N areas therefore the detected regions are basically reduced to the central part of the galaxy. 
The third column of Fig. \ref{fig:HaIS} shows the \OIHa\ line ratio maps for the different velocity bins. The value of these ratios varies between $\sim$0.01 and 0.3.
In our maps, we show in yellow the areas with ratios \OIHa\ $\geq$ 0.1 (or log(\OIHa ) $\geq$ -1.0). Comparing the location of these areas to the \OIIIHa\ maps, shows that these high \OIHa\ areas have \OIIIHa\ values between 0.43 and 1.2 or their equivalent log(\OIIIHb ) between 0.1 and 0.5 assuming case B recombination, with \Ha /\Hb\ $= 2.86$ \citep{Veilleux1987}).
The combination of these line ratios can not be reproduced by photoionization models of HII regions with typical ionization parameters (log(U) between -3.5 and -2) and the SMC-like metallicity of Haro 11 \citep{Kewley2001}. They are located above the HII-star forming area and within the LINERs-shocked gas area in the [OI]-BPT diagram \citep{Veilleux1987,Kewley2001}

This combination of line ratios, however, can be explained by fast radiative shock models. \citet{Allen2008} showed that the value of these line ratios strongly depends on the metal abundance and magnetic parameter of the ISM. In the \texttt{shock + precursor} models, both line ratios increases, although in different manner, with increasing shock velocity. By comparing the observed values of the high \OIHa\ (yellow) areas to the \texttt{shock + precursor} models of \citet{Allen2008} with the Haro 11 metal abundance, we find that this would correspond to shocks with velocities from 200 to 600 \kms .

We also note that the high \OIHa\ areas are consistently associated with areas of low \OIIIHa\ ratios (blue-green areas in the \OIHa\ maps, i.e. at 150 \kms ). This would suggest a relatively low ionization radiation compared to the surrounding gas with much higher \OIIIHa\ ratio. 

Due to the fact that neutral oxygen is found in the neutral to lowly ionized gas, the shocks hinted here likely arise from this gas. In the highly ionized gas, we would need a different line diagnostic to trace shocks as the oxygen atoms would be all ionized (the ionization potential of neutral oxygen is almost identical to that of hydrogen). We will present an extended analysis of shocks, inclusive a more precise prediction of the shock velocities and the excitation mechanisms in Haro 11 in a future paper.

\section{Discussion}

In this section, we discuss the ionized gas assembly of Haro 11 and its connection with the LyC escape. Furthermore, we examine the connection seen between the lowly ionized gas and the diffuse Lyα emission in the galaxy.

\subsection{Ionized gas structures and their origin}

Haro 11 shows complex kinematics of the ionized gas in a broad velocity range. Mechanical feedback appears to be the main driver of the internal kinematics at all velocities, as it has built and accelerate most of the structures seen in our diagnostic maps. The dynamics of the merger however, is traced only in a narrow velocity width at  central and redshifted velocities, as evidenced by the tidal tails in the outermost part of the halo.

\begin{figure}
    \centering
    \includegraphics[width=6cm]{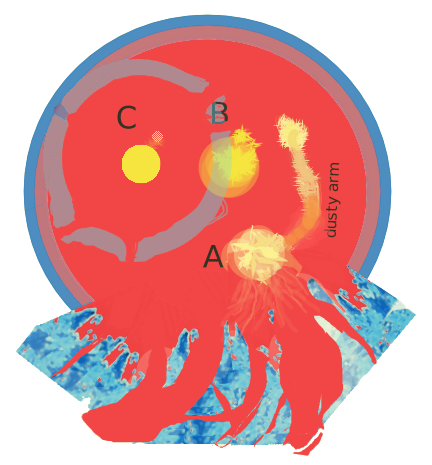}
    \caption{Simplified schematic view of the ionized gas components of Haro 11 as interpreted from our maps. The three star-forming knots A, B, and C are the engines that power the feedback in this galaxy. Energy injections of these sources might have inflated a superbubble ($\sim$ 3.3 kpc), that appears to be open in the south, where the nearly radially oriented filaments originate. Knot C appears to have locally developed a smaller superbubble, whose thick shell is fragmented in several parts. The shell of the largest superbubble might be thin and fully ionized. (Sketch adapted from \citealt{Deharveng2010}, Fig. 4)
    }
    \label{fig:Haro_architecture}
\end{figure}

Fig. \ref{fig:Haro_architecture} shows a sketch of the main ionized gas components that might be found in the centre of Haro 11, as interpreted from our maps. The central star-forming knots A, B, and C and in less proportion the dusty arm are the main sources that produce the strong stellar feedback in the galaxy. They appear to be surrounded
by a superbubble ($\sim$ 3.3 kpc) that is open in the south, where the roughly radial oriented kpc-scale filaments develop. Knot C seems to be surrounded by a thick fragmented supershell ($\sim$ 1.7 kpc) of lowly ionized gas. 
Additionally, but not seen in Fig. \ref{fig:Haro_architecture}, there seems to be a compact kpc-scale structure of lowly ionized gas that extend from the E to the N of knot C. The halo, which might start from the edges of the outer main supershell as inferred from the weak \Ha\ emission, is populated by a filamentary structure towards the S and
three tidal tails towards the E, NE, and N.

Besides, Haro 11 has a large number of secondary structures, such
as small-scale filaments and clumps, which we do not refer here to
keep the simplicity in our analysis. We note however that identifying
well-defined structures in our rich data becomes very hard, hence
the features seen in our maps are subject to interpretation. The
majority of these structures are clearly observed in the \Ha\ maps
and/or ionization maps, while other less prominent structures are
additionally supported by features that hint to their existence. In the
following subsections, we describe these structures in more detail.

\subsubsection{Evidence of kpc-scale superbubbles}

We find evidence of two kpc-scale superbubbles. The main superbubble (r$\sim$ 3.3 kpc) is surrounding the star-forming knots. Principal evidences of its existence come from the bright \Ha\ arc (NE) seen mainly at blueshifted velocities that partially covers a dark cavity, while the halo towards the N is free of structures (at blueshifted velocities) and its \Ha\ intensity decreases strongly with radius. Further evidence comes from the group of filaments that develop roughly in radial direction towards the S. These features are clearly seen at blueshifted velocities and might describe the morphology of the superbubble sketched in Fig. \ref{fig:Haro_architecture}. 

At redshifted velocities, the main supershell might have small openings in the north as hinted by the relatively smaller filaments that emanate. However, the filaments that predominate towards the S, appear to belong to the tidal structures. The second smaller ($\sim$ 1.7 kpc) superbubble is centred at Knot C. Its structure can be seen in the \OIIIHa\ map at v$=-300$ \kms .

Simulations have shown that only a spatio-temporal clustering of SNe can develop kpc-scale superbubbles, whose evolution and fate strongly depend on the SN rate, the mean gas density and the ambient density distribution \citep{Tenorio-Tagle1999,Hopkins2012,Keller2015,Kim2017,Kim2018,Fielding2018}. 
Both superbubbles in this galaxy seem to be very different, not only in the number of SN that might have driven them, their ages and sizes, but principally in their fate and their impact in the galaxy.

The main superbubble might be collectively inflated by SNe from all star clusters located in the centre, mainly within the three central knots. Because a large number of star clusters might be involved, the time between SN events might be short. \citet{Kim2017} showed that for higher SN rate, the energy injected in the ambient medium develops a hot superbubble that rapidly propagates into the ISM. In these superbubbles, a significant fraction of the total injected energy, remains in the bubble while a smaller fraction is radiated away. Thus, this superbubble might have expanded rapidly and at some point, due to Rayleigh--Taylor instabilities, the shell might have fragmented, resulting in the hot material blowout of the bubble. It is not clear which mechanism has favoured the breakout of the bubble in the S. 
The group of filaments towards the S suggest that the shell has fragmented in several places at
the same time, creating several openings where the interior hot gas has vented. \citet{Fielding2018} has demonstrated that a clustering of SNe develops superwinds, powerful enough to escape the galaxy. These galactic winds are able to eject a considerable fraction of gas to the IGM. The authors also showed that the dense ISM clumps struck by the winds are immediately ripped forming a filamentary structure while some fraction might entrain in the wind. 
The same process might have formed the filamentary complex in the S of Haro 11, where part of the shell might be dragged by the superwinds. In the simulations presented \citet{Fielding2018}, the superbubble breaks before the last SN in the cluster has exploded. Thus, the energy released by post-breakout SNe will easily escape the galaxy though the channels cleared by the superwinds.

Typical velocities of galactic winds range from the escape velocity of the galaxy to thousand of \kms\ \citep{Heckman2015}. Assuming that the escape velocity of Haro 11 is about 400 \kms\ (derived in a subsequent subsection), the time needed to develop the filaments of $\sim$10 kpc seen in the S, range from 10 to 25 Myr, for galactic winds with velocities between 1000 and 400 \kms . This might be the time range when the superbubble has fragmented. 
Typical time for superbubble breakout range from few Myr to about 10 Myr \citep{Hopkins2012,Kim2017}. Thus,
the main superbubble might have started to develop between 15 and 35 Myr ago by the current starburst population. Because the cluster formation rate steeply decreases with look-back time in Haro 11, and the number of SNe to develop such superbubble might be very high, it is likely that the main superbubble started to develop about
15 Myr ago.

This superbubble might have had similar physical properties as the superbubble that has developed the bipolar outflow in M82. Both structures seem to have collectively been inflated by SNe of many massive star clusters within their starburst region. In both galaxies, the superbubble breakout might have developed galactic winds with velocities larger than the escape velocity. Additionally, M82 and Haro 11 might have similar escape velocities \citep{Strickland_2000}. 
Thus, the different fate of the superbubble in M82 is exclusively due to the morphological structure of the galaxy.
Superbubbles that develop in disc galaxies rapidly reach a scale height perpendicular to the disc, in both side of the minor axis. The strong density gradient causes a fast acceleration of the bubble and its subsequent quick fragmentation, whose interior hot gas is able to open a wide outflow such as the outflow seen in M82 \citep{Strickland2009,Fielding2018}. In case of Haro 11, the radial density might be somewhat homogeneous. After breakout, the hot gas has then vented from several small openings.

The smaller superbubble centred at Knot C appears to be driven by SNe originated exclusively in this knot. It is covered by a thick shell and filled by highly ionized gas. The shell is already fragmented, and is leaking the hot interior gas to the surrounding regions. This mechanism is supported by two narrow highly ionized channels (Fig.\ref{fig:details} b) that are most likely tracing the path of the outflowing hot gas outwards. Knot C has a dynamical mass of  $\sim$ 10$^8$ \Mo\ \citep{Oestlin2015} and is very compact. It is populated by star clusters in a wide range of ages \citep{Adamo2010}. The circular shape of its shell suggests that it has fragmented when the interior gas was hot and highly overpressure (see e.g. Fig. 4 \citealt{Kim2017}). 
It is not clear when this superbubble has developed, since Knot C seems to have been forming stars continuously over a larger period of time. Nevertheless, it could have been created or most likely strongly accelerated short after its cluster formation peak at $\sim$ 10 Myr \citep{Adamo2010}.
Contrary to the main superbubble, this structure seems to have retained less energy in the bubble, so that at the moment of the blowout it has developed, relative to the main superbubble, weak outflows.

\subsubsection{The merger-driven structures and the lowly ionized gas structure tracing the Ly$\alpha$ emission }

The largest structures assembling the ionized gas of Haro 11 are three tidal tails that are probably lying in a plane perpendicular to
the line of sight, as hinted by their low-velocity dispersions. Along these structures, gas is condensing in clumps. Three small faint
clumps are visible in the farthermost part of the largest tidal tail. An additional closer bright clump is detected in the stellar continuum
in our MUSE data, suggesting that some star cluster are forming in the halo.

\begin{figure}
    \centering
    \includegraphics[width=\columnwidth]{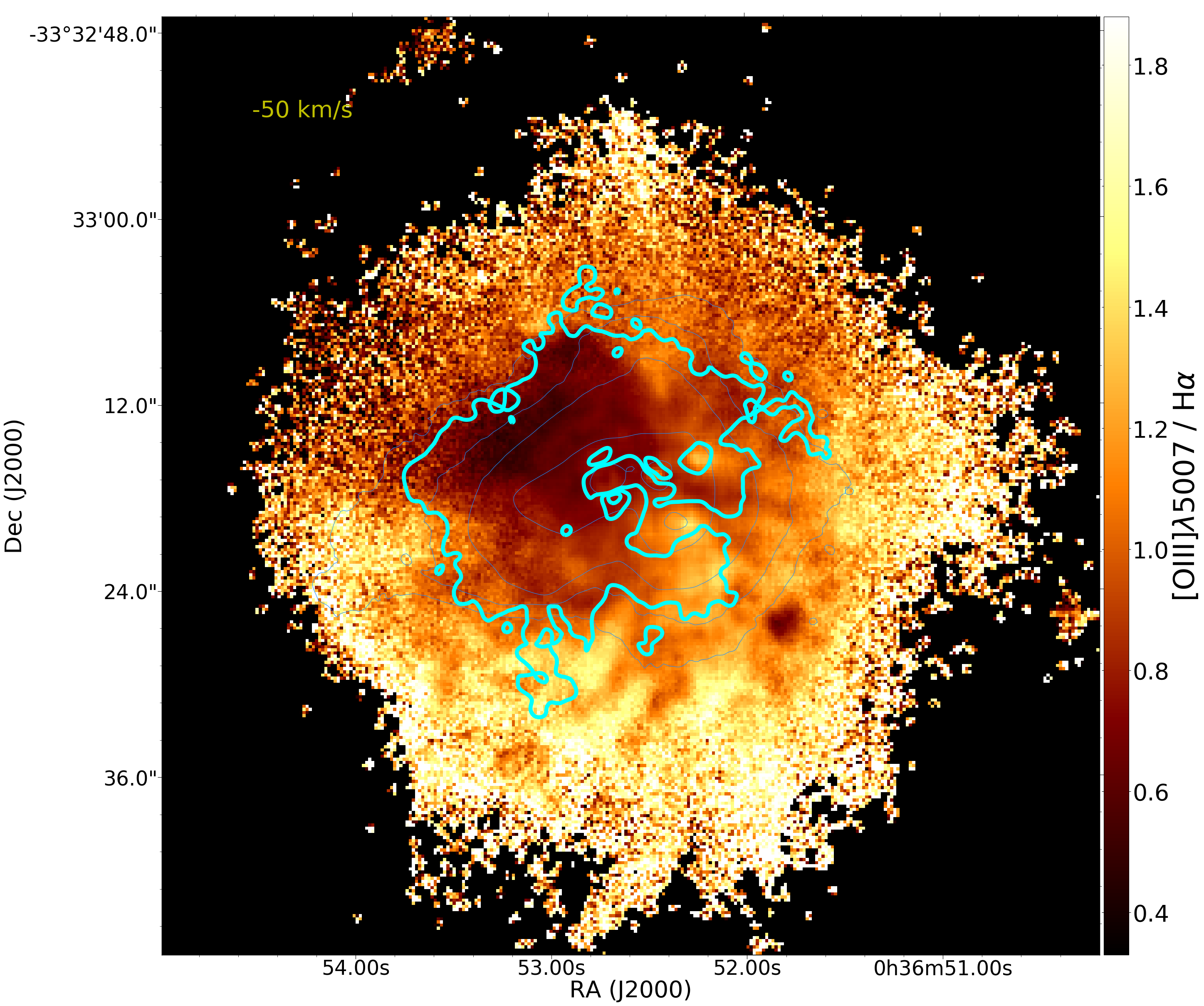}
    \caption{Ionization map at v$=$ -50 \kms\ from Fig. \ref{fig:HaIS}. The principal contour marks the surface brightness of the \La\ flux at 2.0 $\times$ 10$^{-18}$ erg s$^{-1}$ cm$^{-2}$ arcsec$^{-2}$, while contours at 15 and 20 $\times$ 10$^{-18}$ erg s$^{-1}$ cm$^{-2}$ arcsec$^{-2}$ are seen at the bases of knot C, showing the direction of the apparent bipolar outflow. Most of the \La\ flux traced here comes from a diffuse emission that indirectly traces the HI gas. The neutral gas appears to be confined in the lowly ionized structure. }
    \label{fig:Lya_Haro11}
\end{figure}

A huge lowly ionized gas structure is also seen in our maps, that turn from E to N from -250 to 200 \kms . 
This region overlap well with the diffuse Ly$\alpha$ emission (see Fig. \ref{fig:Lya_Haro11}) from \citet{Hayes2007}. This emission originates from resonance scattering of Ly$\alpha$ photons with neutral hydrogen atoms and might suggest that the lowly ionized gas structure is tracing the structure where the neutral HI gas is concentrated. \citet{pardy2016} found that the neutral gas peak at 50 \kms\ and extend in a narrow velocity range at central velocities, although they were unable to map the HI gas distribution in the galaxy.  
Shocks with velocities ranging from 200 to 600 \kms\ are traced in the same area and could have enhanced somehow the diffuse Ly$\alpha$ emission here.

Interestingly, we do not see \La\ emission in the highly ionized area. This suggest that the highly ionized areas might have hydrogen column densities high enough to absorb the \La\ radiation, even in the galactic holes transparent to LyC photons, that could exist in this region.

\subsection{The highly ionized gas and the escape routes of Lyman Continuum radiation}

The ionization structure of Haro 11 shows a fully ionized halo, but is especially enhanced towards the S and W at blue- and redshifted
velocities. Copious amounts of ionizing photons produced mainly in central actively star-forming zone seem to travel further out,
ionizing the halo.

Knot B appears to be the most powerful source of ionizing photons, as it has developed a galactic-scale highly ionized channel towards the N(see Fig.\ref{fig:details}, \textbf{a}) and two towards the W and S conjointly with knot A and the dusty arm. 
However, it is not clear whether the large amount of ionizing photons needed to develop these channels are produced solely by the young stellar population of knot B, or there is a contribution of a black hole in low accretion mode as suggested by \citet{Prestwich2015}. 

Haro 11 is the first local of a handful of galaxies where LyC has directly been measured\citep{Bergvall2006}. 
LyC leakers need to have a highly ionized medium with low column density of neutral gas, at least along the line of sight to the production places for LyC photons to escape.

From the observed starburst-driven LyC leakers, several mechanism have been proposed to favour the escape of LyC radiation, for instance a porous ISM \citep{Puschnig2017}, the presence of galactic holes \citep{Chisholm2015} and a density bounded ISM with  optically thin gas \citep{Vanzella2018}. But up to date, it has been impossible to spatially resolve the origin place of this radiation mainly because of resolution limitation of the UV-detector instrument. Moreover, the mechanisms favouring the escape of LyC radiation in these galaxies have not been studied in detail, as most of them are compact and not well resolved with the current instruments.

In Haro 11, \citet{Hayes2007} and \citet{Rivera-Thorsen2017} suggested LyC to be leaking from knot C, since it has an intense 900\AA\ radiation field and its gas was found to be clumpy. Knot C together with knot B were promoted by \citet{Prestwich2015}, by virtue of luminous soft and hard X-ray radiation respectively. Lately, \citet{Keenan2017} favoured knot A owing to its high ionization values.

Our exceptionally detailed ionization maps shows indeed that the highly ionized zones in Haro 11 extend towards the S-SW at blueshifted velocities, in agreement with \citet{Keenan2017}. Knot A is a favourite LyC leaking candidate, since it hosts several young clusters and is in general highly ionized although at large velocities the ionization level decreases. In our analysis we found evidence of a fragmented superbubble that might have originated powerful galactic winds that in turn may have created galactic holes where LyC photons can escape.

Nevertheless, the density bound scenario might also be an important mechanism in Haro 11.
This scenario supports the escape of LyC photons in the greatest ionized zones in the halo, towards the S and W, but particularly at $\geq$4 kpc south of knot A (see Fig \ref{fig:HaIS}, v$=$-250 \kms ). Thus, ionizing photons might escape in many directions.
In Haro 11, both mechanisms might be at work.

\subsection{Can the gas escape the galaxy?}

\begin{figure*} 
    \centering
    \includegraphics[width=2\columnwidth]{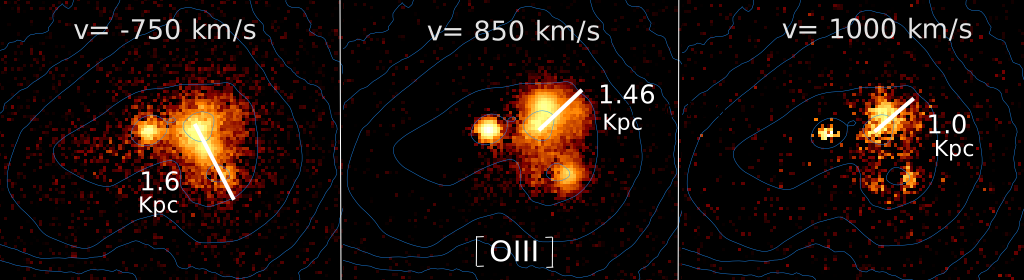}
    \caption{\OIII\ emission gas seen at velocities -750, 850 and 100 \kms\ respectively. The white lines mark the size in kpc of the emitting regions. These are used to calculate the escape velocity of the outflowing gas. Knot B appears to be the main driver of fast outflows.}
    \label{fig:escvel}
\end{figure*}

Fig. \ref{fig:escvel} shows \OIII\ emission maps in 50 \kms\ bins at velocities v$=$ -750, 850 and 1000 \kms .
At higher blueshifted velocities, the [OIII] line is contaminated by a faint emission line ( FeIII $\lambda$4986 ).
The white line in each map shows the size of the outflowing gas, that is used to derive the escape velocities. These are: 1.6 kpc at v$=$-750 \kms , 1.46 kpc at v$=$850 \kms and 1.0 kpc at v$=$1000 \kms.  

To examine if this gas will escape the galaxy, we estimate the escape velocity in Haro 11, following the recipe described in \citet{Marlowe1995}. 
They use a simple approximation of a spherical symmetric isothermal potential and a halo cutoff at $r_{max}$. 
The escape velocity at a radius \textit{r} is then:

\begin{equation}
    v_{esc}= \sqrt[]{2} \cdot v_{circ} \, \sqrt[]{1 + ln (r_{max}/r)} 
\end{equation}

The circular velocity ($v_{circ}$) can be approximated using the virial theorem: v$_{circ}$ $= \sqrt[]{G\cdot M_{tot}/r_{max}}$.
The total mass $M_{tot}$ is estimated from the baryonic matter (M$_{B}$) and the dark matter (M$_{DM}$) : $M_{tot} = M_{B} + M_{DM}$. 
Haro 11 has a stellar mass of 1.6 \x 10$^{10}$ \Mo\ \citep{Oestlin2001,Oestlin2015}. Its ionized gas mass of 1.4 \x\ 10$^9$ \Mo\ (derived here from the total \Ha\ luminosity.) is larger than the neutral gas mass of 5 \x\ 10$^8$ \Mo\ \citep{pardy2016}. The molecular gas mass, however, is not well constrained. It ranges between half to three times the neutral gas as inferred by CO and dust measurements \citep{Cormier2014}.
The baryonic matter is then $M_{B}\simeq 1.9 \times 10^{10}$ \Mo\ , using the highest molecular gas mass derived by \citet{Cormier2014}. 

There is no precise information about the dark matter content in Haro 11.
However, \citet{Oestlin2015} has derived a dynamical mass of  $\sim$ $10^{11}$ \Mo\ assuming that gravitational motions determine the outer velocity field inferred from \Hb\ and [OIII] observation. In this assumption, $\sim$80\% of the total matter is in form of dark matter. 
We consider other two values of dark matter content; we have therefore 3 cases:
(i) 70\%, (ii) 80\% and (iii) 90\% of dark matter.

In addition, the gas seen in Fig. \ref{fig:escvel} could be flowing out in a different direction than the line-of-sight and therefore the intrinsic length of the outflows might be larger for smaller inclinations.
We can not estimate the inclinations ($\alpha_{incl}$) of the outflowing gas with respect to the line-of-sight, 
but we consider three values for $\alpha_{incl}$: 10, 45 and 80.

\begin{table*}
\centering
    \begin{tabular}{l|ccc|ccc|ccc}
    \toprule
    ~    & \multicolumn{3}{c}{\large{\textbf{70\% DM}}} & \multicolumn{3}{c}{\large{\textbf{80\% DM}}},& \multicolumn{3}{c}{\large{\textbf{90\% DM}}} \\
    ~    & ($\alpha_{incl}$ = 10)   & ($\alpha_{incl}$ = 45)   & ($\alpha_{incl}$ = 80)   &  ($\alpha_{incl}$ = 10)   & ($\alpha_{incl}$ = 45)   & ($\alpha_{incl}$ = 80)   & ($\alpha_{incl}$ = 10)   & ($\alpha_{incl}$ = 45)   & ($\alpha_{incl}$ = 80)   \\ \hline
    \large{ \textbf{v$_{esc}$}}    & 238 $\pm$6 & 327 $\pm$6 & 345 $\pm$2 & 291 $\pm$12 & 401 $\pm$14 & 422 $\pm$19 & 412 $\pm$7 & 567 $\pm$9 & 597 $\pm$11 \\
    \large{\textbf{v$_{proj}$}} & 234 $\pm$6 & 231 $\pm$6 & 59 $\pm$0  & 287 $\pm$12 & 283 $\pm$10 & 73 $\pm$3  & 406 $\pm$7 & 401 $\pm$7 & 103 $\pm$2 \\
    \bottomrule 
    \end{tabular}
    \caption{Row 1) Average escape velocities (v$_{esc}$) calculated for different inclination angles and dark matter content for the three outflows seen in Fig \ref{fig:escvel}.  Row 2) Projected escape velocities of these outflows, after correcting them by the inclination effects in the velocity.
    }
    \label{tab:velesc}
\end{table*}

Replacing the values in the equations above, we derived the escape velocities for each combination of the total mass, inclination and outflow lengths. For a given total mass and inclination, we then average the escape velocities calculated for the three outflows seen in Fig \ref{fig:escvel}. The results are presented in the Table \ref{tab:velesc}. The estimated escape velocities (v$_{esc}$) ranges from 240 to 600 \kms , increasing with dark matter content and outflow\textquotesingle s inclination.

The intrinsic velocities of these outflows are at the same time affected but in an opposite manner by the inclinations.  Outflows with higher inclinations are observed at very low velocities compared to their intrinsic velocities.
Table \ref{tab:velesc} shows the line-of-sight projected velocities (v$_{proj}$) for our estimated escape velocities. 
Outflows with inclinations $\alpha_{incl}\geq80$ and velocities of the order of the escape velocities will be detected at velocities lower than 100 \kms\ for dark matter fractions up to 90\%.
However, outflows escaping with inclinations between 10 and 45 will approximately be observed at 230, 285 and 400 \kms , 
for dark matter fractions of 70, 80 and 90\% respectively.

We then derived the ionized gas mass that will escape the galaxy. 
First, the hydrogen ionized gas mass (M$_{H}$)
was calculated from the \Ha\ luminosity (L$_{H\alpha}$), 
assuming an electron density (n$_e$) that decreases exponentially from 4.9 cm$^{-3}$ at 0 \kms\ 
to a lower limit of 1 for v$\geq$ 350 \kms\ as follows:

\begin{equation}
    M_{H,v_ {esc}} = \frac{\mu m_{H} \lambda_{H\alpha} L_{H\alpha,v_{esc}}}{h c \alpha^{eff}_{H\alpha} n_{e}}
\end{equation}

where $\mu$ is the atomic weight and is chosen to be 1, $\lambda_{H\alpha}$ is the wavelength of \Ha\ line , m$_{H}$ is the hydrogen mass, $h$ is the Planck\textquotesingle s constant, c is the speed of light, and $\alpha^{eff}_{H\alpha}$ is the \Ha\ recombination coefficient for case B. 
Then, we have derived the total ionized gas mass assuming that 25\% of total gas mass is in form of helium. Haro 11 has an hydrogen ionized gas mass (M$_{H}$) of $1.0 \times 10^9$ \Mo\ and a total ionized gas mass (M$_{HII}$) of $1.4 \times 10^9$ \Mo . These properties are shown in Table \ref{tab:IonisGasHaro}. 
Lastly, we have consider the ionized gas mass at velocities v$\geq$v$_{esc}$ (L$_{H\alpha ,v_{esc}}$) for the gas that will escape the galaxy.

\begin{table}
    \begin{tabular}{cccc}
    \toprule
    \multicolumn{2}{l}{Total ionized gas mass (M$_{HII}$)}    & \multicolumn{2}{c}{13.7 $\times 10^8$ \Mo }   \\
    \multicolumn{2}{l}{Total hydrogen ionized gas mass}     & \multicolumn{2}{c}{10.3 $\times 10^8$ \Mo }     \\
        ~                               & ~             \\ \hline
        ~                               & ~             \\ 
    \multicolumn{4}{c}{ionized gas mass (M$_{HII, v\geq vel}$) (\Mo )} \\
    $v\geq$100   &  $v\geq$200 & $v\geq$300 &   $v\geq$400 \\ \hline
    9.1 $\times 10^8$    & 5.3 $\times 10^8$ & 2.8 $\times 10^8$ & 1.2 $\times 10^8$ \\ 
    \bottomrule
    \end{tabular}
    \caption{ionized gas properties of Haro 11. Additionally we show the the ionized gas mass for the gas observed at velocities larger than 100, 200, 300 and 400 \kms .} 
    \label{tab:IonisGasHaro}
\end{table}

We consider only the gas that is flowing out at inclinations $\leq$ 45.
For velocities v$\leq$ -450 and v$\geq$ 450 \kms, 
the \Ha\ luminosity was inferred from the \OIII\ surface brightness using a conversion factor chosen as the median value of the galaxy average \OIIIHa\ ratios in the velocity range from -400 to 350 \kms .

\begin{figure}  
    \centering
    \includegraphics[width=8cm]{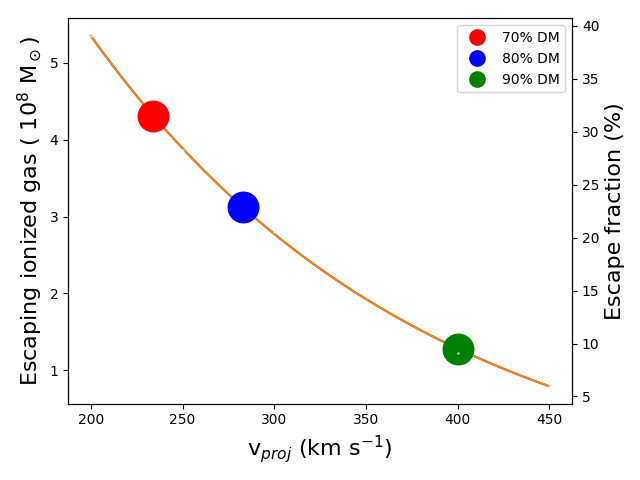}
    \caption{Derived ionized gas mass that will escape the galaxy as function of its line-of-sight escape velocity component.
    The red, blue and green dots are the calculated ionized gas mass that will escape the gravitational potential of Haro 11 for dark matter fractions of 70, 80 and 90\% of the total mass respectively and outflows inclinations $\alpha_{incl}\leq45$.
    }
    \label{fig:fraction_escape}
\end{figure}

The results are shown in the Fig \ref{fig:fraction_escape}. Assuming a dark matter fraction of 70, 80 and 90 \% , the fraction of ionized gas mass that will escape the galaxy is about 31, 23, 9 \% respectively (red, blue and green dot). The orange line shows the ionized gas mass that will escape the galaxy as function of the line-of-sight projected escape velocities.

\section{Conclusions}
Thanks to the unprecedented capability of the MUSE instrument in 
combining the spectral and spatial information at the same time, we were able to study the ionization structure of Haro 11 in maps of 50 \kms\ bins, in a velocity range of -400 to 350 \kms . Here we analyse the impact of stellar feedback by means of the \Ha\ radiation, the state of the ionized gas and the presence of fast shocks in the lowly ionized gas.  In summary:

\begin{itemize}
\item {The ionized gas of Haro 11 is rich in structures that appears to be exclusively shaped by effect of its intense stellar feedback. This includes arcs, bubbles, filaments and galactic ionized channels whose configuration in the ionized gas structure changes with velocity.}
\item {Perhaps the most striking structure uncovered in our maps is the presence of a superbubble ($r\sim 3.3$ kpc), that is already fragmented in the south originating the complex filamentary structure seen at blueshifted velocities. The interior hot gas might have powered superwinds clearing galactic channels where LyC photons can escape. Given that Haro 11 is a LyC leaker, it is very likely that one of these channels is along the line of sight connecting us directly to the LyC source.}
\item {The southwestern hemisphere shows the highest ionization values along the velocity range and has been found to be density bound (\"{O}stlin in prep.) favouring the escape of LyC radiation. Therefore, this mechanism might also allow the escape of LyC photons in Haro 11. }
\item {Knot B appears to be a powerful source of ionizing radiation as in its bases arises galactic highly ionized channels. However it is not clear whether the black hole in low accretion mode suggested by \citet{Prestwich2015} might have contribute to create this structure.}
\item {The stellar feedback of knot C has created a local kpc-scale superbubble ($r\sim 1.7$ kpc) visible at -300 \kms\ . This bubble appears to be already fragmented and its interior hot gas seems to escape the halo through narrow highly ionized channels.
}
\item {The ionization maps reveal a huge lowly ionized gas structure that rotates counterclockwise with velocities. This structure overlaps with the diffuse \La\ emission found by \citet{Hayes2007} which is originated by \La\ photons scattering in the neutral HI gas. Therefore, this structure might trace the location of the neutral gas mixed perhaps with the lowly ionized metal gas. This structure has most likely been shaped by the merger dynamics, compressing the cold neutral and lowly ionized gas in a compact structure of several kpc.  }
\item {Fast shocks (200 -- 500 \kms ) are present in a low to intermediate ionization zones, almost exclusively within the low ionization structure. These shocks contribute to collisionaly excited metal atoms, while ionizing photons might be responsible for the ionization structure.}
\item {Haro 11 shows gas emission at very high velocities. Assuming various dark matter fractions and outflows inclinations, we have derived escape velocities that range from 240 to 600 \kms\ , increasing with dark matter content and outflows inclination.}
\item {We calculate that the fraction of ionized gas mass that will escape the galaxy, is about 31, 23 and 9\% for dark matter fractions of 70, 80 and 90\%.  }

\end{itemize}

\section*{Acknowledgements}

The authors acknowledge financial support from the Swedish Research Council and the Swedish National Space Board. M.H. is Fellow of Knut and Alice Wallenberg Foundation. This research has made use Astropy, a community-developed core Python package for Astronomy \citep{Collaboration2013,Collaboration2018} and APLpy, an
open-source plotting package for Python \citep{Robitaille2012}. This research has made use of NASA Astrophysics Data System Bibliographic Services (ADS)



\bibliographystyle{mnras}
\bibliography{biblio} 

\bsp	
\label{lastpage}
\end{document}